\documentclass[aps,prb,onecolumn,,floatfix]{revtex4-2}

\usepackage{amsmath}
\usepackage{amssymb}
\usepackage{graphicx, float}

\newcommand{\bea}{\begin{eqnarray}}
\newcommand{\eea}{\end{eqnarray}}

\newcommand{\parent}[1]{\left( #1 \right)}
\newcommand{\mean}[1]{\left \langle #1 \right \rangle}

\begin{document}


\title{Fully symmetric nonequilibrium response of stochastic systems}

\author{David Andrieux}

\noaffiliation


\begin{abstract}
We show that the nonequilibrium response of Markovian dynamics can be made fully symmetric, both near and far from equilibrium. This is achieved by varying the affinities along equivalence classes in the space of stochastic dynamics.
\end{abstract}


\maketitle

\section*{Context and objectives}

Nonequilibrium response theory has long been studied near equilibrium, where general results such as the fluctuation-dissipation theorem or the Onsager reciprocity relations hold. 
Far from equilibrium, relations for the response coefficients have been derived as consequences of the fluctuation theorem for currents \cite{AG04}. 

However, these nonequilibrium relations are much more intricate than their near equilibrium counterparts. In particular, they are not fully symmetric and involve combinations of multiple response coefficients.

Recently, the author derived a mapping between nonequilibrium fluctuations and an associated equilibrium dynamics \cite{A12b, A12c}. 
Building on this result, we demonstrate that the nonequilibrium response of currents is symmetric within an equivalence class of stochastic dynamics. 
In addition, the nonequilibrium response coefficients can be expressed in terms of equilibrium correlation functions.
To this end, we show that:
\begin{enumerate}

\item The space of stochastic dynamics can be structured into equivalence classes. Each equivalence class is determined by the composition of an equilibrium dynamics and all posssible affinities
%

\item The equilibrium dynamics of an equivalence class determines the fluctuations of all (nonequilibrium) dynamics in that class

\item Within an equivalence class, the nonequilibrium reponse of currents is symmetric at all orders and for any values of the affinities (i.e., both near and far from equilibrium)
\end{enumerate}

\section{The space of stochastic dynamics is structured into equivalence classes determined by equilibrium dynamics}

We consider a Markov chain characterized by a transition matrix $P = \parent{P_{ij}} \in \mathbb{R}^{N\times N}$ on a finite state space \cite{FN1}.
We assume that the Markov chain is primitive, i.e., there exists an $n_0$ such that $P^{n_0}$ has all positive entries. 

The space of stochastic dynamics can be partitioned into equivalence classes. 
We define the operator
\bea
\bar{P}_{ij} = \sqrt{P_{ij} P_{ji} } 
\eea
and the corresponding {\it equivalence relation} 
\bea
P \sim H \quad \text{if there exists a factor} \ \gamma \ \text{such that} \quad \bar{P} = \gamma  \ \bar{H} \,  .
\label{eqrel}
\eea
This relation is reflexive, symmetric, and transitive. 
It thus forms a partition of the space of stochastic dynamics: every dynamics belongs to one and only one equivalence class.\\

\begin{figure}[H]
\includegraphics[scale=.55]{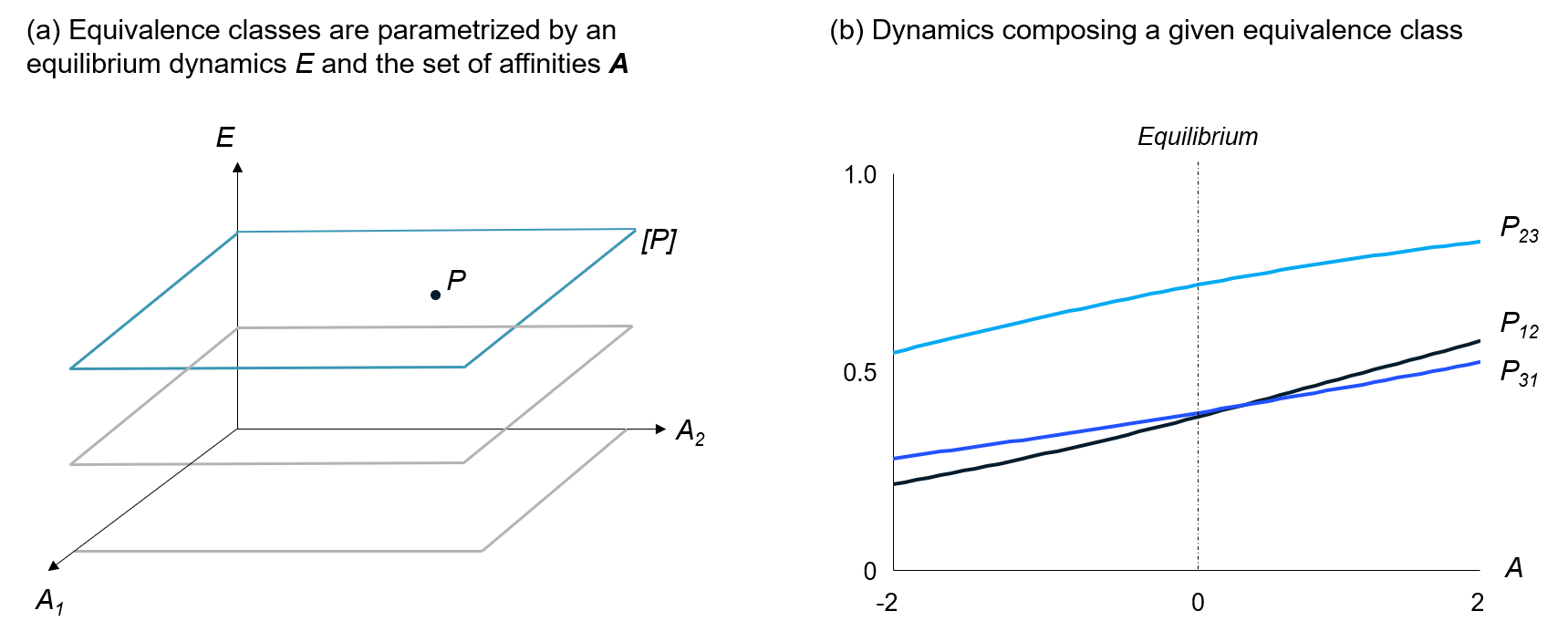}%
\caption{{\bf Equivalence classes and their elements}. (a) The equivalence class $[P]$ is parametrized by the equilibrium dynamics $E[P]$ and the set of all affinities $\pmb{A}$. For simplicity we depict the space of equilbrium dynamics as 1-dimensional.
(b) Dynamics composing an equivalence class typically depend non-linearly on the affinities. As an illustration, we consider a random walk on a ring composed of $3$ sites. The equivalence class is defined by $\bar{E}_{12} = \bar{E}_{21} = 0.2, \bar{E}_{23} = \bar{E}_{32} =0.4, \bar{E}_{31} = \bar{E}_{13} =0.3$. The affinity $A$ then determines each dynamics within $[E]$ and the corresponding transition probabilities $P_{12} = 1-P_{13}, P_{23} = 1-P_{21}, P_{31} = 1-P_{32}$. }
\label{EQTP}
\end{figure}

An equivalence class $[P] = \{ H : H \sim P\}$ is determined by the composition $(E, \pmb{A})$ of an equilibrium dynamics $E$ and the set of all possible affinities $\pmb{A}$ (Fig. \ref{EQTP}, see Ref. \cite{A12b} for a detailed construction).  
In particular, to any $P$ we can associate an equilibrium dynamics $E[P]$ as follows. 
Let $\rho$ denote the Perron root of $\bar{P}$ and $x$ its right Perron eigenvector. If $D = {\rm diag}(x_1, \ldots, x_N)$ we have that
\begin{eqnarray}
E [P] = \frac{1}{\rho} D^{-1} \ \bar{P} \ D
\label{mapT}
\end{eqnarray}
defines an equilibrium stochastic dynamics $E \sim P$. 

\section{The equilibrium dynamics determines the fluctutations of all the nonequilibrium dynamics within its equivalence class}

We consider the thermodynamic currents $\pmb{J} = \{ J_\alpha \}$ and their fluctuations (see Ref. \cite{S76, AG04} for a definition). The current fluctuations for a dynamics $P$ are characterized by the generating function
\bea
Q_P(\pmb{\lambda}) =\lim_{n \rightarrow \infty} \frac{1}{n} \ln \mean{ \exp \parent{\pmb{\lambda} \cdot \sum_{k=1}^n \pmb{J}(k)} }_P  \, .
\eea
All cumulants can then be obtained by successive derivations of $Q$ with respect to the counting parameters $\pmb{\lambda}$:
\bea
Q_P(\pmb{\lambda}) = \sum_{m=1}^{\infty} \frac{1}{m!} K_{\alpha_1 ... \alpha_m}[P] \, \lambda_{\alpha_1} \cdot\cdot\cdot \ \lambda_{\alpha_m} \, ,
\eea
where we sum over repeated indices. 
In particular, the first-order cumulant equals the mean current, $K_{\alpha} = \mean{J_\alpha}$. Note that the cumulants $K_{\alpha_1 ... \alpha_m}$ are symmetric in $(\alpha_1, ..., \alpha_m)$.

We now examine the fluctuations of dynamics within the equivalence class $[P]$. We denote by $Q_{[P]}(\lambda, \pmb{A})$ the generating functions of the elements $(E[P], \pmb{A})$ within $[P]$. We then have the
\\

{\bf Theorem.} {\it Let $[P]$ be an equivalence class and $E[P]$ its associated equilibrium dynamics. Then their current generating functions are related as
\bea
Q_{[P]} (\pmb{\lambda}, \pmb{A}) = Q_E (\pmb{\lambda}+\pmb{A}/2) - Q_E (\pmb{A}/2) \, .
\label{NEQfluct}
\eea
}

DEMONSTRATION: 
In Ref. \cite{A12c}, we showed that the nonequilibrium current fluctuations of an arbitrary nonequilibrium dynamics $P$ are determined by its corresponding equilibrium dynamics $E[P]$ according to 
\bea
Q_{P} (\pmb{\lambda}) = Q_E (\pmb{\lambda}+\pmb{A}/2) - Q_E (\pmb{A}/2) \, .
\label{NEQfluctind}
\eea
By construction, all dynamics in the same equivalence class $P'\sim E$ will satisfy the relation (\ref{NEQfluct}) with the same function $Q_E$. As a result, relation (\ref{NEQfluct}) holds for all dynamics $P' \sim P$ with affinities $\pmb{A}$. Conversely, for any affinity $\pmb{A}$ there exists a corresponding dynamics $P'$ in the equivalence class. The relation (\ref{NEQfluct}) therefore holds for all $\pmb{A}$ within the same equivalence class. $\Box$\\ 

 \begin{figure}[H]
\centering
 \includegraphics[scale=.56]{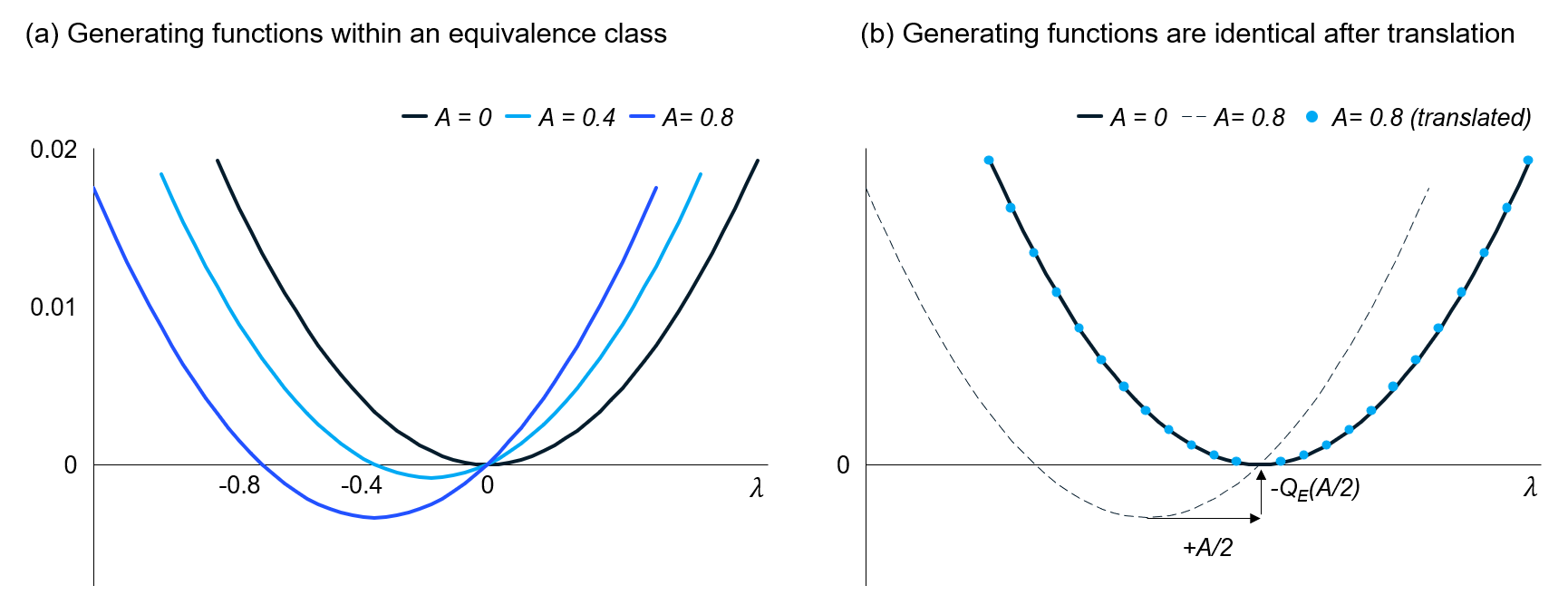}%
 \caption{{\bf Generating functions for different dynamics within an equivalence class.} (a) The generating functions are numerically calculated for different values of the affinity $A$ within the equivalence class. The generating functions obey the fluctuation symmetry $Q(\lambda) = Q(-A-\lambda).$ (b) The generating functions are identical images of each other and related according to (\ref{NEQfluct}). All parameters are identical to Fig. \ref{EQTP}.}
\label{Qshift}
 \end{figure}

In other words, the generating function of the current $Q_P$ is entirely determined by the equilibrium fluctuations $Q_E$ (Fig \ref{Qshift}) 
\cite{FN3}. 
Therefore, $E[P]$ acts as a reference equilibrium dynamics that is intrinsic to $P$. 
We also note that the generating functions $Q_{P}$ is directly related to the factor $\gamma$ defining the equivalence relation (\ref{eqrel}). If $\bar{P} = \gamma \bar{E}$ then $Q_P =-2 \log \gamma$ \cite{A12b}.

\section{The nonequilibrium response is symmetric within an equivalence class and expressed in terms of equilibrium correlations}

\begin{figure}
\centering
\includegraphics[scale=.59]{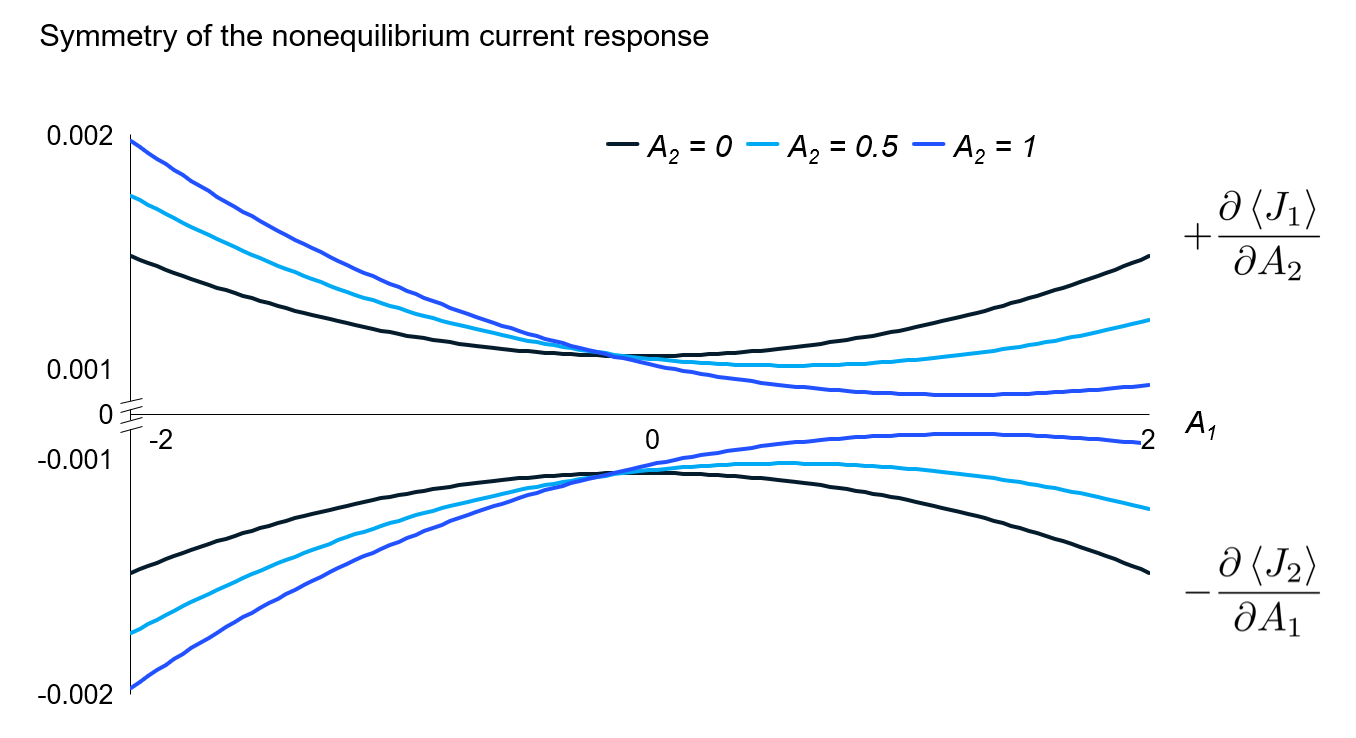}%
\caption{{\bf Symmetry of the nonequilbirum current response}. We consider a $4$-state system defined by the equivalence class $\bar{E}_{1,2} = \bar{E}_{2,1} = 0.2, \bar{E}_{2,3} = \bar{E}_{3,2} =0.4, \bar{E}_{3,1} = \bar{E}_{1,3} =0.5, \bar{E}_{3,4} = \bar{E}_{4,3} =0.1, \bar{E}_{4,1} = \bar{E}_{1,4} =0.1$. This system has $2$ independent currents and affinities, here measured along the cycles $(1,2,3)$ and $(1,3,4)$. The symmetry $\partial J_1/ \partial A_2 = \partial J_2/ \partial A_1$ is verified for multiple values of the affinities (for visual clarity, we depict $-\partial J_2/ \partial A_1$ and verify that it's the mirror image of $\partial J_1/ \partial A_2$).}
\label{NLRsym}
\end{figure}

We now explore the consequences of Eq. (\ref{NEQfluct}) on the response theory. 
As the right-hand side of Eq. (\ref{NEQfluct}) only depends on the affinities through the translation $\pmb{\lambda}+\pmb{A}/2$, the first order response of the current $\mean{J_\alpha} = K_\alpha (\pmb{A})$ reads
\bea
\frac{\partial \mean{J_\alpha}}{\partial A_{\beta} } (\pmb{A}) =  \frac{1}{2} \frac{\partial^2 Q_E}{\partial \lambda_\alpha \partial \lambda_\beta} (\pmb{A}/2) = \frac{\partial \mean{J_\beta}}{\partial A_{\alpha} } (\pmb{A})\, 
\label{NLR1}
\eea
which is symmetric in ($\alpha, \beta$) for any $\pmb{A}$, i.e. both near and far from equilibrium (Fig. \ref{NLRsym}). 
Also, since $Q_E(\pmb{\lambda}) = Q_E(-\pmb{\lambda})$, the current is antisymmetric with respect to $\pmb{A}$, $\mean{J_\alpha} (-\pmb{A}) = -\mean{J_\alpha} (\pmb{A})$, while its response $\partial \mean{J_\alpha}/ \partial A_{\beta}$ is symmetric.

More generally, the response of any cumulant is expressed as
\bea
\frac{ \partial^l K_{\alpha_1 ... \alpha_m}(\pmb{A}) } {\partial A_{\beta_1} ... \partial A_{\beta_l} } 
= \parent{\frac{1}{2}}^l \frac{\partial^{m+l} Q_E  }{\partial \lambda_{\beta_1} ... \partial \lambda_{\beta_l}  \partial \lambda_{\alpha_1} ... \partial \lambda_{\alpha_m} } (\pmb{A}/2) \, , 
\label{NLRfull}
\eea
which is fully symmetric in ($\alpha_1, ... , \alpha_m, \beta_1, ..., \beta_l$) for all $\pmb{A}$. The cumulant response is symmetric in $\pmb{A}$ if $m+l$ is even, and antisymmetric otherwise.
Alternatively, expression (\ref{NLRfull}) reveals that the current response and cumulants are directly related:
\bea
K_{\alpha_1 ... \alpha_m}(\pmb{A}) = 2^{m-1} \, \frac{ \partial^{m-1}\mean{J_{\alpha_1}}(\pmb{A}) } {\partial A_{\alpha_2} ... \partial A_{\alpha_m} } \, .
\eea
In other words, knowledge of the currents $\mean{J_{\alpha}}(\pmb{A})$ is sufficient to obtain all cumulants.

We recover the traditional response theory by expanding the currents as functions of the affinities around $\pmb{A} =0$:
\bea
\mean{J_\alpha} = \sum_{l=1}^{\infty} \frac{1}{l!} L_{\alpha, \beta_1 ... \beta_l} \, A_{\beta_1} \cdot\cdot\cdot \ A_{\beta_l} \, ,
\eea
where we sum over repeated indices. Using Eq. (\ref{NLRfull}) and the symmetry $Q_E(\pmb{\lambda}) = Q_E(-\pmb{\lambda})$, the response coefficients can be expressed in terms of equilibrium correlations:
\bea
L_{\alpha, \beta_1 ... \beta_l} &=& 0  \quad  \quad \quad \quad \quad \quad \quad  \quad \text{if {\it l} is even} \label{L.a} \\
L_{\alpha, \beta_1 ... \beta_l} &=& \parent{\frac{1}{2}}^l K_{\alpha \beta_1 ... \beta_l} (\pmb{0}) \quad \text{if {\it l} is odd.}
\label{L.b}
\eea 
This shows that all response coefficients $L_{\alpha, \beta_1 ... \beta_l}$ are fully symmetric in ($\alpha, \beta_1, ..., \beta_l$). 
In particular, we recover the Onsager symmetry $L_{\alpha, \beta} = L_{\beta, \alpha}$ and the corresponding Green-Kubo formula, $L_{\alpha, \beta} = (1/2) K_{\alpha\beta} (\pmb{0}).$ 
Similar expressions can be derived for the higher-order cumulants and their nonequilibrium response \cite{FN4}.

Remarkably, expressions (\ref{L.a}) and (\ref{L.b}) take a simpler form than the ones derived in Ref. \cite{AG04}, which involve combinations of multiple cumulants and their nonequilibrium responses. Here the coefficients either vanish or can be expressed as equilibrium correlations \cite{FN5}.

\section*{Discussion}

\begin{figure}[h]
\includegraphics[scale=.62]{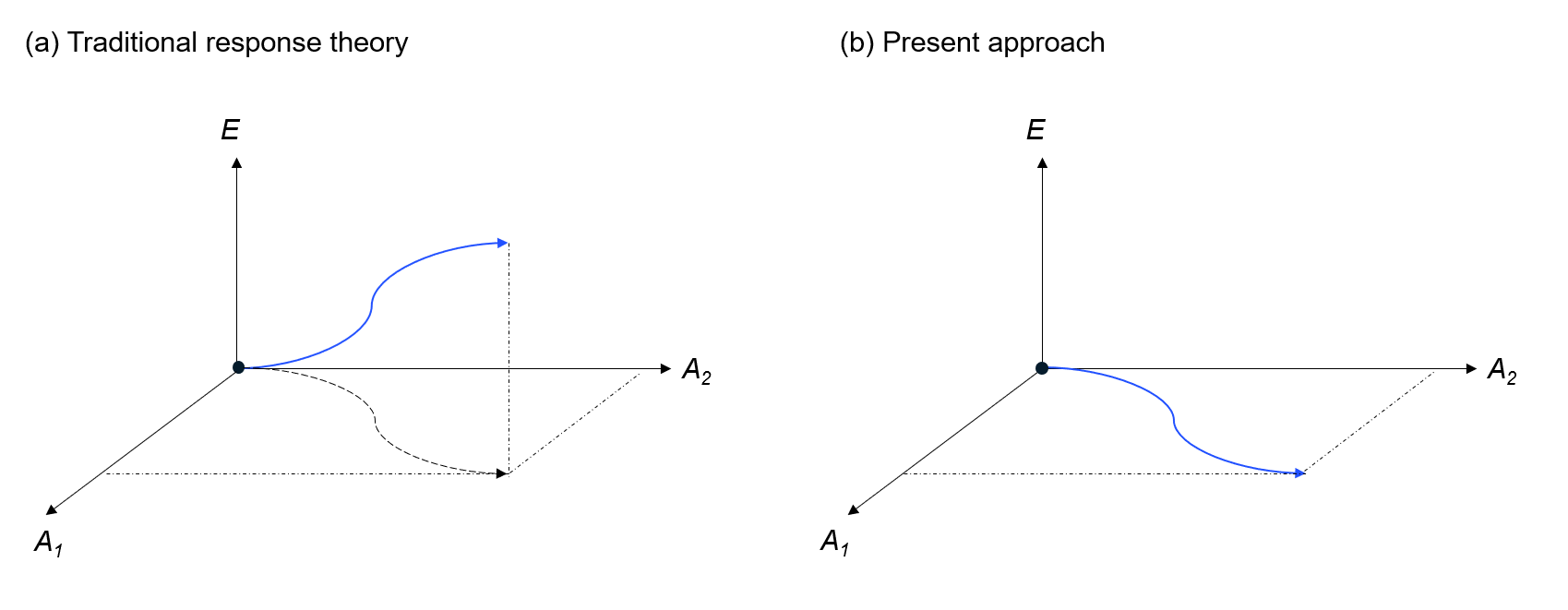}%
\caption{{\bf Schematic comparison of the traditional and the present approach to the response theory}. In both cases, the system traces the same path in the space of affinities. (a) In the traditional response theory, both the affinities and equivalence classes change. (b) In the present approach, the system remains in its original equivalence class when varying the affinities.}
\label{Difftheory}
\end{figure}

The results (\ref{NLR1}) and (\ref{NLRfull}) complement the traditional response theory. 
In the traditional response theory, the dynamics shifts across equivalence classes as the affinities are varied (Fig. \ref{Difftheory}a).
As a result, the shape of the generating function changes as a function of the affinities in addition to being translated. 
This leads to additional contributions of the form $\partial K/ \partial A_\alpha ... \partial A_\beta$ in the response theory, breaking the symmetry (\ref{NLRfull}).
In contrast, here the dynamics remains in its original equivalence class when varying the affinities (Fig. \ref{Difftheory}b). 
This leads to symmetric response coefficients that can be expressed as equilibrium correlation functions.

Note that varying affinities while staying in a given equivalence class typically requires changing multiple transition probabilities at the same time (Fig \ref{EQTP}b). 
As experimental setups improve their ability to manipulate mesoscopic degrees of freedom, it will become possible to use equivalence classes to control the nonequilbrium transport properties of fluctuating systems.
In any case, the present results offer a new way to explore nonequilibrium systems by leveraging their intrinsic dynamical characteristics.


\vskip 1 cm

{\bf Disclaimer.} This paper is not intended for journal publication.


\begin{thebibliography}{99}
\bibitem{FN3} The fluctuation theorem $Q(\pmb{\lambda}) = Q(-\pmb{A}-\pmb{\lambda})$ is satisfied as $Q_E (\pmb{\lambda}) = Q_E (-\pmb{\lambda})$ for an equilibrium dynamics
\bibitem{FN4} Expanding the cumulants $K_{\alpha_1 ... \alpha_m} (\pmb{A}) = \sum_l (1/l!) L_{\alpha_1 ... \alpha_m, \beta_1 ... \beta_l} A_{\beta_1} \cdots A_{\beta_l}$, a direct calculation from Eq. (\ref{NLRfull}) shows that $L_{\alpha_1 ... \alpha_m, \beta_1 ... \beta_l} = (1/2)^l K_{\alpha_1 ... \alpha_m \beta_1 ... \beta_l} (\pmb{0})$ if $m+l$ is even and $0$ otherwise.
\end{thebibliography}
\end{document}